# Soft matter mechanics of immune cell aggregates

S. Askari,[1,†] G. Saldo Rubio,[2,†] A. Datar,[1] H. Harjunpää,[2] S. C. Fagerholm,[2,‡] and M. Backholm[1,*]

[1]Department of Applied Physics, Aalto University, Espoo, Finland
[2]Research Program of Molecular and Integrative Biosciences, Faculty of Bio- and Environmental Sciences, University of Helsinki, Helsinki, Finland
[†]Equal contribution

T-cells are a crucial subset of white blood cells that play a central role in the immune system. When T-cells bind antigens, it leads to cell activation and the induction of an immune response. If T-cells are activated by antigens *in vivo* or artificially *in vitro,* they form multicellular aggregates. The mechanical properties of such clusters provide valuable information on different T-cell activation pathways. Furthermore, the aggregate mechanics capture how T-cells are affected by mechanical forces and interact within larger conglomerates, such as lymph nodes and tumours. However, an understanding of collective T-cell adhesion and mechanics following cell activation is currently lacking. Probing the mechanics of fragile and microscopically small living samples is experimentally challenging. Here, the micropipette force sensor technique was used to stretch T-cell aggregates and directly measure their Young's modulus and ultimate tensile strength. A mechanistic model was developed to correlate how the stiffness of the mesoscale multicellular aggregate emerges from the mechanical response of the individual microscopic cells within the cluster. We show how the aggregate elasticity is affected by different activators and relate this to different activation pathways in the cells. Our soft matter mechanics study of multicellular T-cell aggregates contributes to our understanding of the biology behind immune cell activation.

## I. INTRODUCTION

Cells are the fundamental building blocks of life. In multicellular organisms, individual cells work collectively to maintain the structural and functional integrity of tissues and organs. To study these multicellular living materials with the rigour of biological physics and mathematics, multicellular aggregates have been proposed as the ideal model system (*1*). Decoding how cells behave and interact in an aggregate is key to understanding diverse biological processes, from tissue regeneration (*2*) and morphogenesis (*3*), to immune responses (*4*) and the progression of diseases like cancer (*5*).

Microscale force probes, such as atomic force microscopy, optical tweezers, and micropipette-based techniques, have been widely used to measure the mechanical properties of single cells (*6–13*) as well as their adhesion to surfaces or other cells (*14–23*). In addition, non-direct force measurements have been performed using, for example, the centrifugation assay, spinning disk, and flow chamber, to probe the adhesion force of populations of single cells (*24*). However, probing the biomechanics and dynamics of multicellular aggregates is challenging due to the mesoscale length scale of the system. This has led to the development of several new force probes and experimental approaches. For example, Foty *et al*. used a parallel plate compression apparatus to probe the interfacial tension of liver cell aggregates (*25*). This approach was later adapted by several others (*26*, *27*). Kalantarian *et al*. used centrifugation to measure the surface tension of aggregates (*28*). Ryan *et al*. probed the spreading rates of fibroblast cell aggregates to probe their cell-cell cohesion (*29*) and the group of Brochard-Wyart has further studied the spreading and flow of cellular aggregates (*30–32*). In the same group, Guevorkian *et al*. also developed the micropipette aspiration technique to directly measure the viscoelastic mechanical properties of cancer cell aggregates (*33*) as well as the mechanosensitive "shivering" of the aggregates under controlled aspiration (*34*). Furthermore, Gonzalez-Rodriguez *et al*. measured the detachment and fracture of cellular aggregates using a force-calibrated glass slide as a cantilever (*35*). As a final example, Lyu *et al*. have used a soft resistive force-sensing diaphragm to probe the tiny contractile force of cardiac organoids (*36*). In general, experimental soft matter physics has been instrumental in forming a new understanding of these living multicellular materials.

From a soft matter mechanics perspective, T-cells are a particularly interesting cell type. These highly specialized, microscopically small immune cells are crucial in the adaptive response of the immune system of mammals (*37*). T-cells originate in the bone marrow and migrate to the thymus for further development and maturation. The mature T-cells circulate in the blood stream, encounter antigens in lymph nodes, become activated, and finally migrate to peripheral tissues to participate in the intricate functions of the whole immune system. Their main task is to recognise and destroy foreign invaders such as bacteria and viruses. They do this by recognising and binding to a specific structure called an antigen on, for example, bacteria. Seminal biophysical studies have shown interesting mechanisms of how T-cells use mechanical forces in their immunological response (*38*, *39*).

[‡]susanna.fagerholm@helsinki.fi
*matilda.backholm@aalto.fi

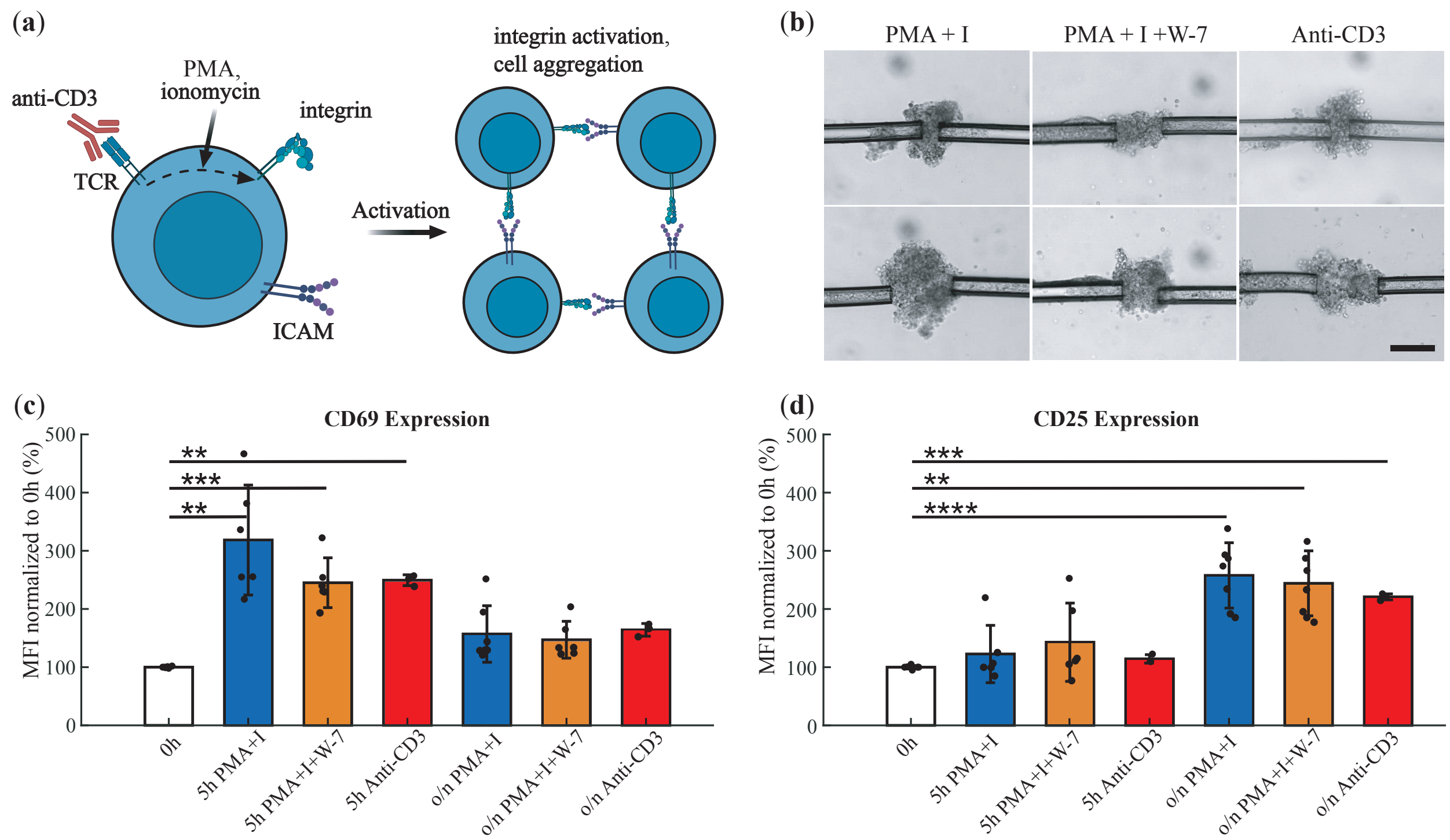

**FIG. 1. *In vitro* activation of T-cells. (a)** Overview of *in vitro* activators: Anti-CD3 binds the T-cell receptors (TCRs) and triggers the downstream activation signaling. PMA and ionomycin together mimic part of this signaling cascade. W-7, a calmodulin inhibitor, interferes with some effects of calcium influx into T-cells. T-cell activation leads to integrin activation, integrin binding to intercellular adhesion molecules (ICAMs), and cellular aggregation. It also induces various other cellular changes, including the expression of T-cell activation markers (not shown). **(b)** Optical microscopy images of T-cell aggregates (held by two micropipettes) following activation with PMA and ionomycin (PMA+I, left), PMA, ionomycin, and W-7 (PMA+I+W-7, middle), and anti-CD3 (right). Scale bar 50 µm. **(c)** Expression levels of the early T-cell activation marker CD69 after stimulation with PMA + ionomycin, PMA + ionomycin +W-7, or anti-CD3 at 5 hours post-induction or overnight (o/n). Presented as MFI (mean fluorescence intensity) as measured using a flow cytometer. **(d)** Expression levels of the late T-cell activation marker CD25 under the same conditions. The stars in (c)-(d) indicate statistical significance of Welch's t-test of the mean of a pair of groups with $^{**}p < 0.01$, $^{***}p < 0.001$, $^{****}p < 0.0001$.

Upon encountering a specific antigen, T-cells undergo activation and clonal expansion, that is, a rapid division to produce many identical cells that can respond to the same antigen (*37*). To study the biology of T-cell activation, the activation process can be mimicked in an antigen-agnostic manner. This means that the T-cells are activated in a way that does not depend on recognizing a specific antigen but still triggers their activation and expansion. Such artificial activation can be done using chemicals such as phorbol myristate acetate (PMA) and ionomycin (*40*), or anti-CD3 antibodies (*4*, *22*). These different activators work through distinct mechanisms (**FIG. 1a**). Anti-CD3 binds directly to the T-cell receptor (TCR), mimicking antigen binding and triggering a signalling cascade that results in T-cell activation. PMA and ionomycin stimulate elements of the intracellular signalling pathway that follow antigen recognition. Another important compound for the T-cell activation pathway is the inhibitor W-7, which inhibits the function of calmodulin ($Ca^{2+}$-binding protein) in cells (*41*).

When T-cells are activated, this leads to integrin activation and T-cell clustering into large aggregates of many cells (**FIG. 1b**). These aggregates are biologically important, as they allow the cells to efficiently share important soluble factors (interleukin-2) that allow for cell proliferation (*42*). Integrins play a vital role in the immune cell function, particularly in T-cell adhesion, mediated by interactions with intercellular adhesion molecules (ICAMs) (*21*). Atomic force microscopy studies have revealed time-dependent adhesion strengthening, with unbinding forces increasing from 140 to 580 pN over short time scales (*21*) as well as the effect of force generation on T-cell activation (*43*). Advances in optical tweezers have provided further insights into integrin mechanics, enabling the quantification of cell-cell bond rupture forces *(22)*.

These biophysical approaches collectively enhance our understanding of integrin function at molecular and cellular levels. However, an understanding of collective T-cell adhesion and mechanical forces in cell aggregates following cell activation is currently completely lacking. Such information is crucial because immune cells often interact within larger conglomerates, such as lymph nodes and tumours, and mechanical forces between cells are emerging as important regulators of immune cell programming, behaviour, and function (*44*).

Here, we have used a soft matter mechanics approach to take a first step towards understanding the biology of T-cell interactions in mesoscale multicellular aggregates. In our experiments, we have activated mouse primary T-cells to study how the different activation pathways affect the resulting immune cell activity and multicellular aggregate properties. By using the micropipette force sensor (MFS) technique, we have directly measured the mechanical properties of the *in vitro* activated immune cell aggregates through stretching. To correlate the emergent macroscopic mechanical response of the multicellular system with the microscopic cell elasticity, we have developed a new mechanistic model. We have explored the effects of various factors, such as post-induction time, aggregate volume, and pre-stretching, on the T-cell aggregate. Finally, we evaluated the influence of specific activators, including W-7 (a calmodulin antagonist) and anti-CD3, on aggregate stiffness. Our results showcase the importance of a multidisciplinary approach in the study of biological systems.

## II. METHODS

In this section, we concisely describe our experimental approach. For all technical details on methodology and materials, refer to the *Experimental details* section at the end of the article.

### A. *In vitro* induced T-cell aggregation

To study the cohesion and mechanics of immune cell aggregates, primary T-cells were isolated from mice and activated using PMA and ionomycin, or by only anti-CD3. When the calmodulin inhibitor W-7 was used, it was added 30 minutes before the addition of PMA and ionomycin. These chemicals trigger intracellular signalling events that activate integrins, making the T-cells highly adhesive and causing them to aggregate into clusters (**FIG. 1b**). As shown in **FIG. 1c-d,** these treatments induce T-cell activation, as evidenced by increased CD69 expression at 5 hours and CD25 expression after overnight incubation *(45)*. Of these *in vitro* activators, anti-CD3 most closely mimics the biological activation of T-cells, initiating intracellular signalling within the T-cell. PMA and ionomycin mimic these signalling effects by artificially activating the same pathways as T-cell receptor activation (*46*). W-7, a calmodulin antagonist, further modulates T-cell signalling and introduces an additional regulatory dimension to PMA + ionomycin-induced activation (*47*, *48*).

### B. Stretching of T-cell aggregates

The MFS technique (*49*) was used to stretch the T-cell aggregates and measure their mechanical properties (**FIG. 2**). In this technique, the deflection $x$ of a glass micropipette is used to measure and apply a force $F = k_p x$, where $k_p$ is the spring constant of the cantilever, determined through calibration. The MFS technique has been extensively used to probe the biomechanical properties of single cells (*8*, *9*), including T-cells (*39*), as well as microbial flocs (*50*, *51*) and whole organisms (*52*, *53*).

The buffer solution containing T-cells was carefully injected between two glass slides, which were mounted in a custom-made holder on an inverted microscope. Two micropipettes were required for the stretching experiments: one straight holding micropipette mounted on a linear motor and one L-shaped, force-calibrated MFS mounted at a 90-degree angle on a manual *xyz*-micromanipulator (**FIG. 2a**). These micropipettes were moved into the chamber, ensuring that the entire lengths of the cantilevers were immersed in the buffer solution. Both micropipettes were used to hold on to an aggregate with gentle suction using syringes. To stretch the aggregate, the straight micropipette was moved to the left at a constant speed of $v = 20$ µm/s using a linear motor (**FIG. 2b**). This movement caused a deflection in the L-shaped micropipette, which applied an increasing force onto the aggregate, leading to its stretching.

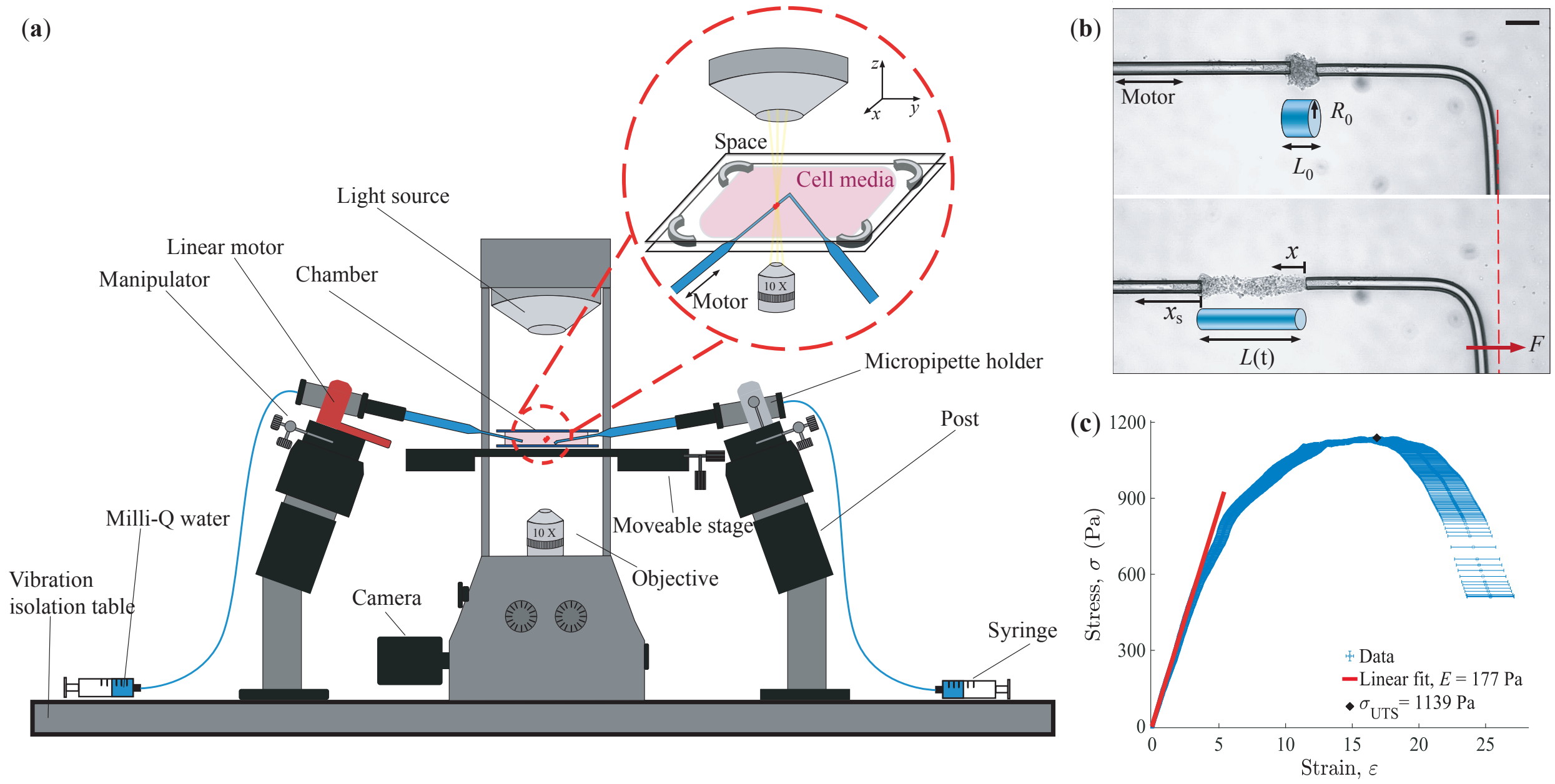

**FIG. 2. Cell aggregate stretching using the micropipette force sensor. (a)** Schematic sketch (not to scale) of the setup with the straight and L-shaped micropipettes holding on to the T-cell aggregate with suction. The straight micropipette is connected to a linear motor and the L-shaped is calibrated and used as a force sensor. **(b)** Examples of optical microscopy images from before (top) and after (bottom) the stretching. The aggregate is modelled as a cylinder with an initial radius $R_0$ and length $L_0$. During the experiment, the straight micropipette is moved to the left at a constant speed ($x_S = vt$), causing the L-shaped force sensor to deflect ($x$). This applies a Hookean force $F = k_p x$ onto the aggregate, which stretches, $L(t)$. Scale bar 50 µm. **(c)** Example of stress-strain graph from a typical stretching experiment. The Young's modulus $E$ and ultimate tensile strength $\sigma_{UTS}$ are determined from the graph. The error bars are error propagations of the stress and strain using the standard deviation of $k_p$, $R_0$ and $L_0$ from several measurements.

## III. MODELS

### A. Mechanical properties

The T-cell aggregate was modelled as a cylinder with an initial radius $R_0$ and length $L_0$ (**FIG. 2b**). When stretched using MFS, the aggregate is under a stress $\sigma = F/\pi R_0^2 = k_p x/\pi R_0^2$ and strain $\varepsilon = \Delta L/L_0$, where $\Delta L = x_S - x = vt - x$ is the change in length of the aggregate. The Young's modulus is defined as $E = \sigma/\varepsilon$ for small deformations within the elastic regime (*54*). This corresponds to the slope of the initial, linear regime in the standard stress-strain graph, illustrated with an example of our experimental data in **FIG. 2c**. As the aggregate is stretched further, the system transitions into the plastic regime, where material deformations become irreversible. From the stress-strain graph, the ultimate tensile strength $\sigma_{UTS}$ was determined (**FIG. 2c**), which is an important measure of the mechanical properties of the aggregate, representing the maximum stress the material can sustain before breaking.

### B. Mechanistic model

Various theoretical models, such as the geometry-based vertex model and lattice-based Cellular Potts model (*1*), exist for describing a multicellular system. Most of these focus on epithelial morphology, topology, dynamics, and mechanics in 2D (*3*, *55–57*), A limited focus has also been on modelling 3D systems (*57–62*). In our case, an advantage of a vertex model over the Cellular Potts model would be that it allows for the treatment of the elastic response of a multicellular system under external loading (*1*). However, the vertex-based model is not suitable for describing T-cell aggregates, where the randomly organized small cells do not deform into polyhedral shapes as required in a 3D vertex geometry. To bridge the gap between individual cellular mechanical response and the emergent macroscopic mechanical properties (defined in Sect. IIIA), a new scaling law-based mechanistic model was developed (**FIG. 3**). A T-cell aggregate was assumed to consists of randomly arranged cells adhered to each other. The mechanical response of the cell (radius $r_c$) was modelled with an elastic spring with a stiffness of $k_c$. This effective cell stiffness includes the stiffness of the cell-cell adhesion bonds coupled with that of the cell itself. The deformations of these springs contribute to the emergent mechanical response of the aggregate.

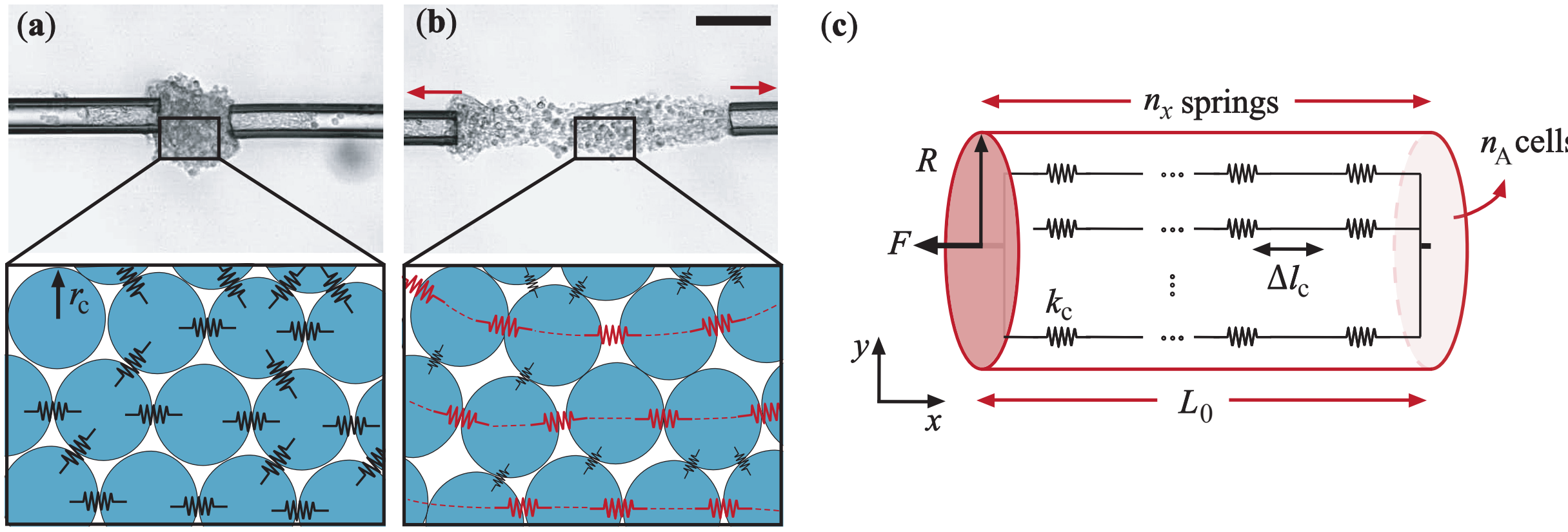

**FIG. 3. Mechanistic model.** The cells are modeled as spheres with a radius of $r_c$ and spring constant $k_c$. The aggregates with their modelled cell-spring structure are shown in **(a)** before and **(b)** during the stretching. Scale bar 50 µm. **(c)** Schematic drawing of the mechanistic model with parallel connections of series-connected springs acting as force chains through the aggregate. Each spring changes its length by $\Delta l_c$ when the external force $F$ is applied.

The force is applied in the *x*-direction, causing a mechanical response of the cells mostly along this axis. In response, each spring extends by $\Delta l_c$, forming a force chain in the aggregate, where neighbouring cells along the *x*-axis create a series connection of springs (**FIG. 3b–c**). For one such series connection along the entire aggregate length, the number of springs is $n_x \sim L_0/2r_c$. To stretch one such series-connected chain of springs, a force of $F_{\text{series}} = k_c \Delta l_c$ would be required (*63*). The change in length of an entire chain upon stretching is the same as the change of length of the entire aggregate: $\Delta L \sim n_x \Delta l_c \sim (L_0/2r_c)\Delta l_c$.

The total number of series-connected chains within the cylindrical cross-sectional area ($A = \pi R_0^2$) of the aggregate is $n_A \sim R_0^2/r_c^2$. These chains of springs respond in parallel when the entire aggregate is stretched (**FIG. 3c**), requiring a total force of $F = n_A F_{\text{series}}$, according to the conventional rules of stretching parallel connected springs (*64*). This gives an expression for the total force needed to stretch the aggregate: $F = k_c \Delta l_c\, R_0^2/r_c^2$. Adding all of this together, the Young's modulus can thus be written as

$$E = \frac{\sigma}{\varepsilon} = \frac{F}{A}\frac{L_0}{\Delta L} \sim \frac{k_c}{r_c}. \tag{1}$$

This equation links the emergent mechanical response of the entire multicellular aggregate to the stiffness of the individual cells and their internal bonds. It shows that the elasticity of the aggregate scales linearly with the elasticity of the cells. This finding is critical for understanding the relationship between T-cell activation levels and aggregate mechanics, as higher T-cell activation can be hypothesised to create stronger adhesion within the aggregate as well as affect the cell stiffness.

## III. RESULTS & DISCUSSION

### A. Mechanics of T-cell aggregates

T-cell aggregates formed through activation with PMA and ionomycin were used as a model system in this study. The cells were activated in a biology lab and the aggregates were then transported to a physics lab for measurements. The activated T-cells remained viable for at least 24h post-induction. An example of a stress-strain curve from an MFS stretching experiment is shown in **FIG. 2c**. The MFS technique is well suited for these measurements because of its ability to precisely measure nN-range forces while simultaneously stretching the sample and observing the resulting deformation. Since the sample is held through gentle suction with the micropipettes, no potentially harmful glue is needed. Several aggregates of varying volumes ($V = 10^4$ to $7 \times 10^5$ µm$^3$) were stretched. In **FIG. 4a**, the Young's modulus is plotted as a function of time from when the activation was done. There is no effect of post-activation time on the stiffness of the clusters – as long as the cells remain viable, the Young's modulus remains constant.

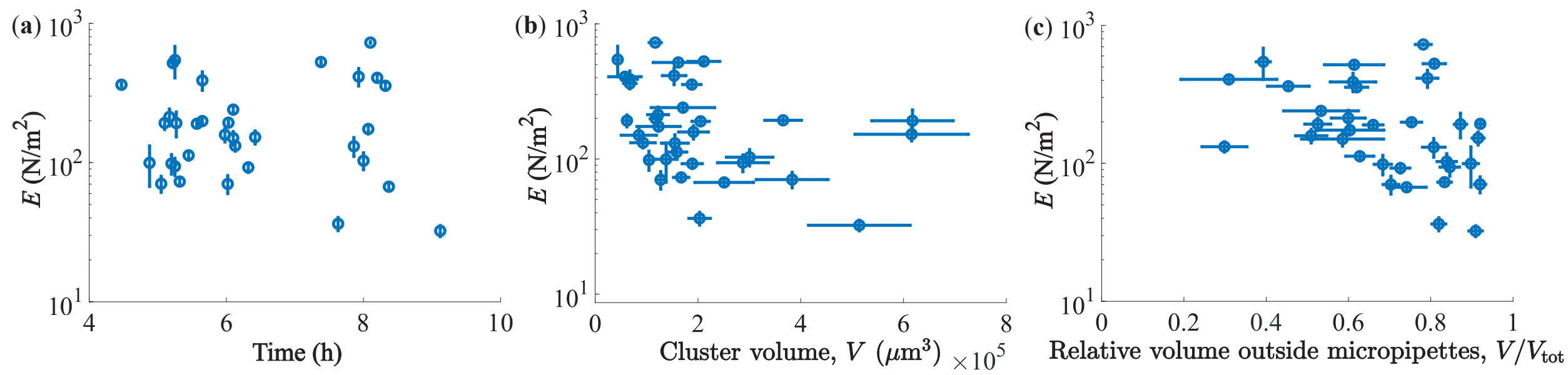

**FIG. 4. Effect of time, volume, and aspiration.** Young's modulus of T-cell aggregates (activated with PMA and ionomycin) as a function of **(a)** time after activation, **(b)** cluster volume, and **(c)** relative cluster volume outside of the micropipettes. Within error, the stiffness remains constant for all cases.

Within error, the measured mechanical properties of the aggregates were independent of their volume (**FIG. 4b**). Experiments were initially also performed on aggregates with volumes much larger than $10^6$ μm$^3$. However, these were found to be more irregular in shape with portions of the cluster not contributing to the stretching process. Since the cylindrical shape approximation did not properly account for the actual stretching of the large aggregates, these clusters were discarded from our results. To hold on to the aggregate during the stretching experiment, a fraction of it was aspirated into both micropipettes, as seen clearly in the optical microscopy images of **FIG. 3**. In **FIG. 4c**, the measured Young's modulus is plotted as a function of the volume $V$ of the aggregate (outside of the micropipettes) divided by the total volume $V_{\rm tot}$ of the aggregate (inside and outside of the micropipettes). Within experimental error, there is no mechanobiological effect of the aspiration on the measured stiffness.

The Young's modulus and ultimate tensile strength data are summarised in **FIG. 5.** The average values across 35 aggregates were determined as $E = 248 \pm 234$ Pa and $\sigma_{\rm UTS} = 390 \pm 334$ Pa (median values shown with black lines in FIG. 5). In comparison, the stiffness of *individual*, non-activated T-cells has been measured as 80–100 Pa (*65*, *66*), whereas activated T-cells have stiffnesses in the range of 100–300 Pa (*39*). The increase in stiffness during activation is due to drastic modifications of the actomyosin cytoskeleton. Our measured mechanical properties are lower than those reported for murine sarcoma aggregates (*33*, *35*). Specifically, micropipette aspiration experiments on these yielded a Young's modulus of $E = 700 \pm 100$ Pa (*33*). Measurements using glass slide-based cantilevers showed Young's modulus values ranging from $E = 1000 \pm 300$ Pa at very low speeds ($v < 0.02$ $\mu$m/s) to $E = 9000 \pm 2000$ Pa at higher speeds ($v = 1$ to $50$ $\mu$m/s) (*35*). Our T-cell aggregates may be softer due to the big differences in adhesion mechanisms and cytoskeletal architecture between immune cells and sarcoma cells. The latter study reported maximum stress values of $\sigma_{\rm max} = 300 - 700$ Pa, which are similar, within error, to our measurements.

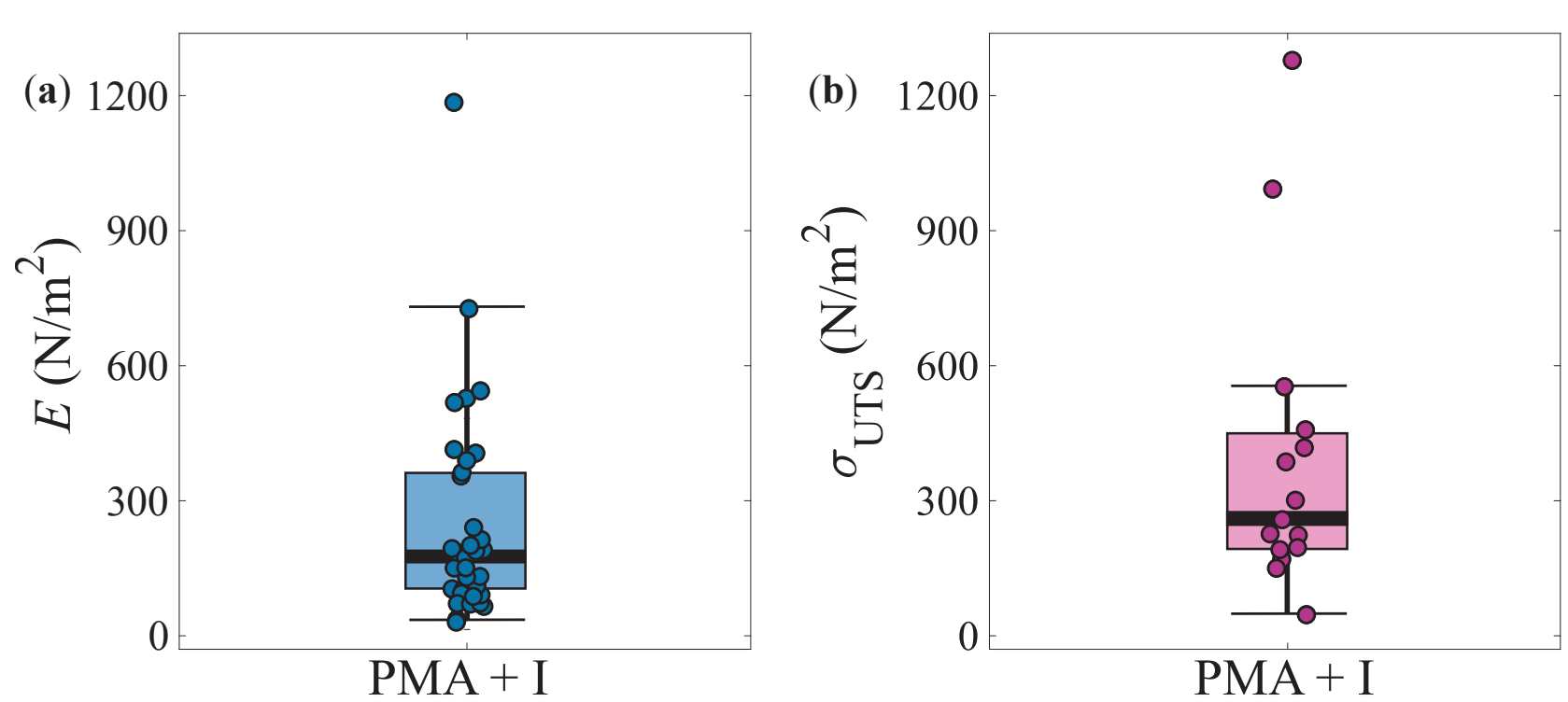

**FIG. 5. Mechanical properties. (a)** The Young's modulus and **(b)** ultimate tensile strength of T-cell aggregates activated with PMA and ionomycin (I). The thick black line inside the box represents the median. The box spans the interquartile range (IQR), covering the middle 50% of the data. The error bars (whiskers) extend to the minimum and maximum values within 1.5 times the IQR, while points beyond this range are considered outliers.

By combining the measured elastic modulus with our mechanistic model (Eq. 1), the elastic response of one cell (including both the stretching of the cell as well as the deformation of the cell-cell bonds) $k_\mathrm{c} \sim r_\mathrm{c}E$ can be determined. The average T-cell radius was measured as $r_\mathrm{c} = 3.3 \pm 0.8\,\mu\mathrm{m}$, giving an average cell spring constant of $k_\mathrm{c} \approx 8 \times 10^{-4}$ N/m. This spring constant agrees well with the results in Lek *et al.* (*21*), where atomic force microscopy was used to measure the adhesion force between a single effector T-cell and an integrin ligand-coated surface. From their data, a spring constant of approximately was extracted $10^{-4}$ N/m, which captures both the deformation of the cell as well as the cell-cell adhesion bonds. This agreement underscores the reliability of our experimental and theoretical approach in quantifying T-cell mechanics within cellular aggregates.

Stretching experiments were also performed on cells containing a mutation in the β2-integrin that affects T-cell adhesion under flow conditions (*67*). These knock-in cells were activated with PMA and ionomycin, but no difference was detected between the knock-in and the wild-type cellular aggregates (see Appendix A). This indicates that the integrin mutation is not important for regulating cell-cell aggregate properties or activation under these activation conditions, as also shown in our previous work (*67*).

### B. Pre-stretching of T-cell aggregates

Cells respond to mechanical stimuli by altering their properties. To investigate this in our aggregates, double-stretching experiments were performed. The aggregates were first pre-stretched to a maximum strain of $\varepsilon_1 \approx 0.2$ to 1, then returned to their initial configurations before being stretched again until rupture. The Young's modulus was measured in both elastic stretching phases, and the change in stiffness ($E_2/E_1$, where 1 and 2 refer to the first and second stretch, respectively) is plotted as a function of $\varepsilon_1$ in **FIG. 6**.

Our results show that aggregates soften in response to pre-stretching by a factor of $0.7 \pm 0.3$. Additionally, there is a slight trend toward stronger softening when aggregates undergo higher pre-stretch strains. This is in contrast to previous findings, where mechanical stimulation led to stiffening due to integrin-ICAM catch bond formation (*68*). Instead, our results suggest plastic deformations within the aggregates during the first stretching phase, potentially due to irreversible bond breakage. Alternatively, pre-stretching may alter the overall structure and shape of the aggregate, influencing its mechanical response.

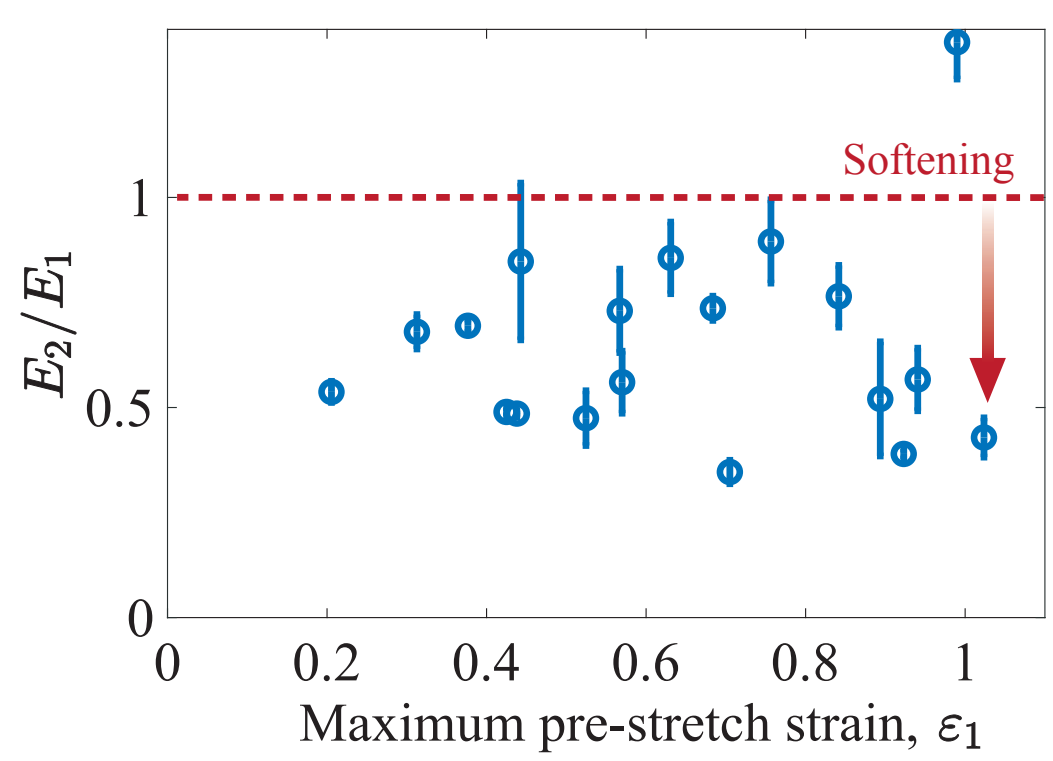

**FIG. 6. Effect of pre-stretching.** The ratio between the Young's modulus in the second ($E_2$) and first ($E_1$) stretching experiment as a function of maximum strain $\varepsilon_1$ during the pre-stretching. The aggregates soften after the pre-stretching.

### C. Effect of T-cell activation pathway

When T-cells were treated with W-7 in addition to the standard PMA and ionomycin, the aggregates exhibited increased stiffness compared to those activated with only PMA and ionomycin (**FIG. 7a**). This effect was more pronounced when comparing data from experiments conducted on the same day, using cells from the same mouse. Specifically, the addition of W-7 increased aggregate stiffness by a factor of $E_\mathrm{PMA+I+W7}/E_\mathrm{PMA+I} = 2.8 \pm 0.7$ (based on same-mouse experiments). This increase directly translates to a stiffening of the cells (including the cell-cell adhesion bonds) by the same factor (Eq. 1). Biologically, this suggests that T-cells become stiffer in response to W-7, likely due to its inhibition of $Ca^{2+}$-calmodulin, which itself is an inhibitor of caldesmon-actin bundle binding. By this mechanism, W-7 may be responsible for enhancing T-cell stiffness (*69*). Interestingly, we also detected a small (not statistically significant) increase in metabolic activity in W-7 treated cells (see Appendix B), indicating that altered mechanical responses of the aggregates may also be reflected in the activation response of the cells, for example altered cell proliferation, which we have not explored here.

To further explore the effect of different activation pathways, T-cells were activated using either anti-CD3 antibodies or PMA and ionomycin. Anti-CD3 antibodies provide a more physiologically relevant activation method, as they induce the same conformational change in CD3 (the T-cell antigen receptor) that occurs when it recognizes an antigen *in vivo*. This conformational change initiates the intracellular signalling cascade. The activation with anti-CD3 rendered a clear decrease in the aggregate stiffness as compared to cells activated with PMA and ionomycin (**FIG. 7b**). Comparing the change in stiffness for aggregates with cells from the same mouse, $E_{\mathrm{Anti-CD3}}/E_{\mathrm{PMA+I}} = 0.30 \pm 0.02$. This suggests a lower level of activation of the T-cells with the antibody as compared to using the more artificial chemicals (which cause maximal activation of the cells). This correlates with lower CD69 expression in anti-CD3 treated cells compared to PMA/ionomycin activation, e.g., lower level of activation (**Fig. 1C**). Anti-CD3 is known to stiffen individual T-cells by a factor of two upon activation (*39*). Our results indicate that PMA + ionomycin either stiffen the cells by a factor of ca 4 upon activation, or that the aggregate stiffness is primarily determined by cell-cell adhesion strength rather than individual cell elasticity.

The immune system is a complex cellular system that mediates functions such as protection against infection and tumor surveillance, but is also involved in pathological events, for example autoimmunity. To mediate these functions, immune cells interact both with other immune cells and non-immune cells, forming interconnected cellular communities. The human immune system consists of approximately 1.8 trillion cells, with a total weight of 1.2 kg. Of these, up to 40% are lymphocytes (T-cells and B-cells) (*70*). These cellular communities interact in a highly coordinated manner for the immune system to perform its functions. However, lymphocytes are small and motile and mediate transient interactions with other immune cells and non-immune cells, making their analysis more challenging than the stable interactions found in other tissues and organs in the body.

Transient interactions between immune cells are mediated by cell surface proteins, and their complexity is beginning to be unraveled (*71*, *72*). Immune cells communicate with each other through these cell surface receptors, which initiate intracellular signaling events within the cells, leading to cellular outcomes such as activation and effector functions. The intracellular signals consist of biochemical signals, but more recently, it has become apparent that also mechanical signals are important to determine outcome in immune reactions, such as lymphocyte activation (*73*). Indeed, increased stiffness of the environment leads to stronger T-cell activation (*74*, *75*). However, mechanical forces in larger immune cell communities have so far remained uncharacterized due to technical challenges to measure such forces in the small and transient cell groups. Here, we have made a first attempt to characterize mechanical forces in immune cell communities, starting with small T lymphocyte aggregates, and to correlate the mechanical forces with cellular activation outcomes, by utilizing novel methodology.

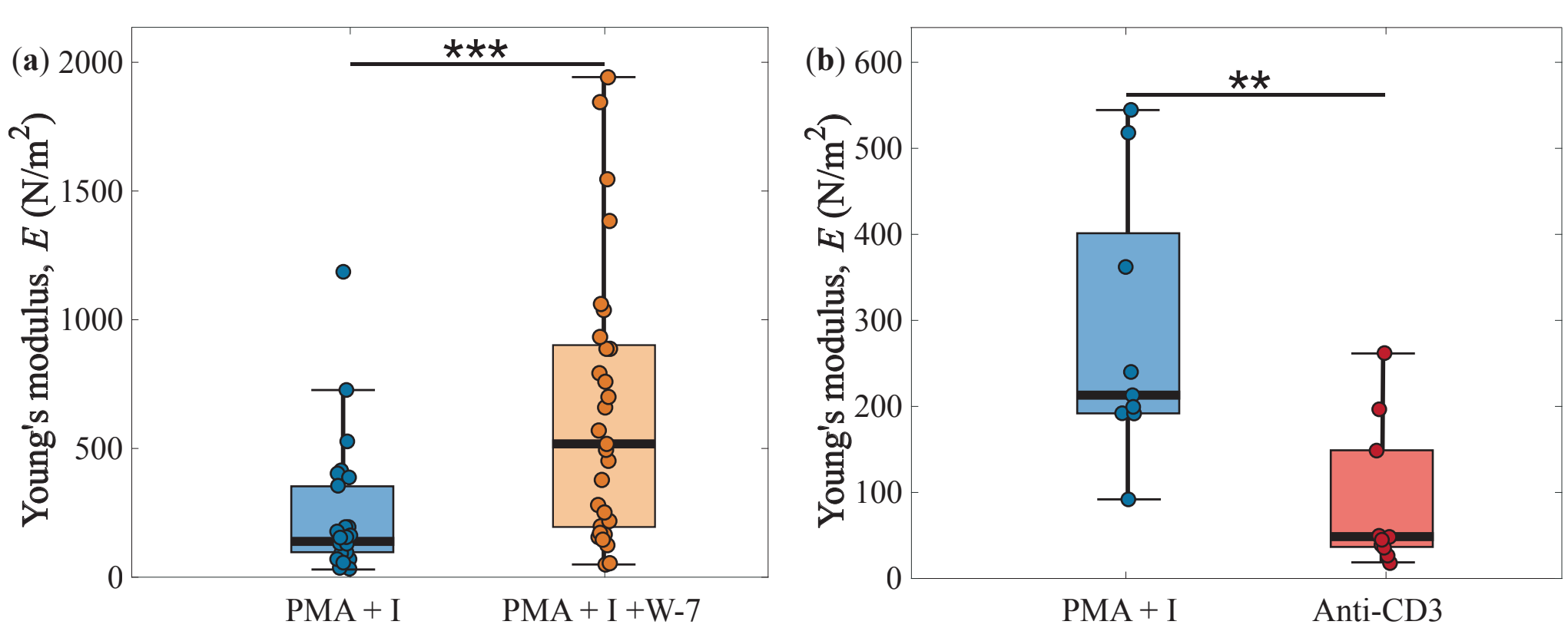

**FIG. 7. Activation pathways.** The Young's modulus of T-cell aggregates activated with PMA + ionomycin compared to **(a)** aggregates activated also with W-7 and **(b)** only with anti-CD3. The addition of W-7 makes the aggregates stiffer, whereas anti-CD3 produces softer aggregates than PMA + ionomycin. The stars indicate $^{**}p < 0.01$ and $^{***}p < 0.001$.

## IV. CONCLUSIONS

In this work the micropipette force sensor technique was used to directly measure the mechanical properties of mesoscale (diameter $\sim 30 - 120\,\mu\text{m}$) T-cell aggregates through stretching experiments. The multicellular aggregates were formed upon *in vitro* activation using PMA and ionomycin and their Young's modulus and ultimate tensile strength were determined as $E = 248 \pm 234\ \text{N/m}^2$ and $\sigma_{\text{UTS}} = 390 \pm 334\ \text{N/m}^2$. These values are similar to the mechanical properties reported in the literature for aggregates made of other cell types. A mechanistic model was developed to couple the emergent mechanical response of the aggregate (Young's modulus $E$) with the intrinsic mechanical response of the cells (spring constant $k_c$): $E \sim k_c/r_c$, where $r_c$ is the average cell radius. This rendered $k_c \approx 8 \times 10^{-4}$ N/m which agrees well with literature values from atomic force microscopy measurements on single T-cells. The addition of W-7 rendered stiffening of the aggregates by a factor of $2.8 \pm 0.7$, likely due to its inhibition of $Ca^{2+}$-calmodulin. Conversely, T-cell activation with anti-CD3 rendered softening of the clusters by a factor of $0.30 \pm 0.02$ compared to aggregates activated with PMA and ionomycin. This suggests a lower level of activation of the T-cells with the antibody as compared to using the more artificial chemicals, which cause stronger activation of the cells.

## V. EXPERIMENTAL DETAILS

### A. T-cell isolation

The day before an extraction the MACS buffer was prepared following Miltenyi's instructions: PBS (Lonza, cat. # 17-516F), 0.5% BSA (Biowest, cat. # P6154) and 2 mM EDTA. After the BSA was fully dissolved, the solution was sterilized through filtration through a 22 µm nylon mesh filter (Fisher Scientific, cat. # 2263548).

Spleens from C57BL/6N mice were extracted and mechanically dissociated by grinding the tissue between the back of a sterile syringe and a 70 µm cell strainer (Fisher Scientific, cat. #22363548) placed over a 50 ml Falcon after the cell strainer was wetted using 1 mL of ice-cold PBS (Lonza, cat. # 17-516F) supplemented with 2% FCS (Gibco, cat # 10500–064). After the tissue was dissociated, the syringe back and the strainer were rinsed using a further 5 mL of ice-cold PBS+2% FCS. The cell strainer was discarded, and the cell suspension was centrifuged at 330g for 5 minutes. The supernatant was discarded, and the pellet was resuspended in room temperature ACK lysis buffer (0.15 mol/L $NH_4Cl$, 0.01 mol/L $KHCO_3$, and 0.1 mmol/L EDTA in milli-Q water) to perform red blood cell lysis. With our aliquot the time needed to do so was 50s. After this time the cell suspension was diluted to a final volume of 20 mL with PBS+2% FCS. The cell suspension was passed through a 70 µm cell strainer and into a new falcon. The suspension was then centrifuged at 330g for 5 minutes and the supernatant discarded. The pellet was resuspended in 40 µL of de-gassed MACS buffer + 10 µL of the Biotinylated Antibody Cocktail from the CD4+ T-cell enrichment kit (Miltenyi, cat. # 130-104-454) and placed on ice for 5 min. After the initial incubation on ice, 30 µL of MACS buffer and 20 µL of the anti-biotin magnetic beads from the same kit were added, mixed well and incubated on ice for 10 min.

During the incubation, the magnetic separation column (Miltenyi, cat. # 130-042-401) was placed on the magnet over a 15 mL falcon and capped with a 70 µm cell strainer and primed by adding 3 ml of MACS buffer. After the 10-minute incubation of the magnetic beads was done, the cell suspension and beads were resuspended in 1 mL of MACS buffer and transferred to the magnetic separation column. After the cell suspension flows through the column, the falcon is rinsed with 1 mL of MACS buffer and the liquid was added to the column. After this, the column was washed with more MACS buffer until 10 ml of MACS buffer and CD4+ T-cell enriched cell suspension made it through the column. The cell suspension was then counted and centrifuged at 330g for 5 minutes. The supernatant was discarded, and the cell pellet was resuspended in RPMI (Lonza, cat. # 12-167F/EuroClone, cat. # ECB9006L) supplemented with 10% FCS (Gibco, cat # 10500–064), 100 U/ml penicillin–streptomycin (penicillin, Orion, cat. # 465161; streptomycin Thermo Fisher Scientific, cat. # D7253-100 g), 2 mM L-glutamine (Thermo Fisher Scientific, cat. # BP379–100) and 50 µM 2-mercaptoethanol (Fluka biochemika, cat. # 63690) at a concentration of 4 million cells/ml and plated in a 24 well plate (CellStar, cat. # 662160) and incubated at 37°C before proceeding to the different experiments.

### B. T-cell treatments

The T-cell suspension was activated using anti-CD3 2,5 µg/mL (Sigma-Aldrich, cat. # MAB484) or PMA 10 ng/mL (Sigma-Aldrich, cat. # P81389) + Ionomycin 0,5 µg/mL (Thermo Fisher Scientific, cat. # J62448.M). In the cases where the calmodulin inhibitor W-7 (Sigma-Aldrich, cat. # 681629) was used, it was added at 50 µM 30 minutes before the addition of PMA + Ionomycin.

### C. Flow cytometry

The following conjugated antibodies were used for flow cytometric analysis of: Puromycin-PE (BioLegend, cat. # 381503, clone 2A4). CD4-PE-Cy7 (BioLegend, cat. # 100528, clone RM4–5), CD25-APC-Cy7 (BioLegend, cat. # 102026, clone PC61), CD69-APC (BioLegend, cat. #104514, clone H1.2F3. Fc-receptor block (BD Pharmingen, cat. # 553142 clone 2.4G2) was used in all experiments assessing mouse cells, and unstained and FMO controls were included in all panels. 7-AAD (eBioscience, cat. # 00-6993-50) was used to detect dead cells. Acquisition was performed on an LSR Fortessa flow cytometer (Becton Dickinson), and data were analyzed using FlowJo software (Tree Star).

### D. Translation rate analysis

Puromycin (Gibco, cat. #A11138-03) is incorporated into nascent proteins in a manner that prevents the further extension of said proteins and assayed through flow cytometry using an anti-Puromycin PE conjugated antibody (BioLegend, cat. #381503) after fixing and permeabilizing the cells using the Foxp3 permeabilization kit (eBiocience, cat. #00-5523-00). To determine the total background for puro-PE fluorescence we used Harringtonine (Sigma-Aldrich, cat. #SML1091-10MG) at 2µg/ml for 30 minutes before the addition of Puromycin. To determine the maximum energy level of the cells, DMSO (Fisher Scientific, cat. # BP231-1) 0.08% was used as a negative control and added for 30 minutes before the addition of Puromycin. To determine the contribution of mitochondrial energy pathways to the total, the mitochondrial ATP synthase inhibitor Oligomycin (Sigma-Aldrich, cat. #04876-5MG) was used at 1µM for 30 minutes before the addition of Puromycin. To determine the contribution of glucose-related energy production to the total, the phosphoglucoisomerase inhibitor 2-Deoxy-D-Glucose (Sigma-Aldrich, cat. #D8375-5G) was used for 30 minutes before the addition of Puromycin. To determine the contribution of glycolysis alone to the cell's energy level a combination of Oligomycin and 2-Deoxy-D-Glucose can be used. In all cases, Puromycin was added at 10 µg/ml for 45 minutes after 30 minutes of incubation with the different inhibitors. The incorporation of puromycin was then assayed via flow cytometry.

### E. Micropipette force sensor

The MFS was manufactured and calibrated as described in detail in (*49*). Briefly, the MFS was made of a hollow glass capillary (World Precision Instruments, TW100-6) with an inner diameter of 0.75 mm and an outer diameter of 1 mm. The cantilever was pulled using a micropipette puller (Narishige, PN-31) and shaped with a microforge (Narishige, MF-900).

Prior to use, the MFS requires calibration to determine its spring constant, $k_\mathrm{p}$. We calibrate our cantilevers using the water droplet method developed by Colbert *et al*.(*9*) and described in detail in(*49*), using water drops of different sizes as control weights. The cantilever was mounted horizontally, and a droplet of Milli-Q water (Millipore Direct-Q 3 UV) was pushed out to rest on the cantilever end using a syringe connected with a plastic tube to the micropipette holder (IM-H3 Injection Holder Set by Narishige Lifemed Co., Ltd.). A camera used to image the experiment from the side at a frame rate of 30 fps. Using an in-house MATLAB (MathWorks) image analysis script, the shape of the drop was modelled as an ellipsoid with a volume of $V$. The drop weight was then calculated as $W = \rho g V$, where $\rho$ is the density of water. The cantilever deflection $\Delta x$ was measured as a function of time through image analysis in MATLAB, and the micropipette spring constant was determined using Hooke's law: $k_\mathrm{p} = W/\Delta x$. The calibration was repeated several times for each cantilever. In this study, micropipettes with spring constants ranging from 10 to 40 nN/µm were used.

### F. Stretching experiments

***Microscope setup***

An inverted wide-field microscope (Nikon ECLIPSE Ts2-FL) equipped with a manually adjustable *xy*-stage was used to securely position the sample chamber. A 10x objective (Nikon 10x/0.25 Ph1 DL Objective Lens) was used to ensure sufficient resolution as well as a big enough field of view to observe the cluster deformation and micropipette motion. Image acquisition was performed using a FLIR camera (GS3-U3-23S6M-C, Integrated Imaging Solutions, Inc.) mounted on the microscope. An image sequence of the experiment was recorded at 30 fps, a frame rate chosen to match the experimental speed and ensure adequate temporal resolution.

***Experimental chamber***

The sample chamber was constructed using aluminium and designed to hold two untreated glass slides at a fixed spacing. Depending on the volume of the cell solution media, two sizes of glass slides were used (Plain Pre-Cleaned Microscope Slides, 75 mm × 50 mm, thickness of 0.96 mm to 1.06 mm, and VWR Ground Edge Microscope Slides, 75 mm × 25 mm, thickness of 0.8 mm to 1.2 mm). Spacing between the slides was maintained using four plastic spacers, each 2 mm thick, placed at the corners of the slides to create a uniform 2 mm gap (**FIG. 2A**). The cell culture media was injected into the chamber using a plastic pipette, with the tip widened by cutting it with scissors to minimize disruption of the clusters during injection. Surface tension prevented leakage from the chamber, while strong intercellular bonds ensured the integrity of the clusters during their transfer from a 24-well plate to the sample chamber.

***Micropipette setup***

Cantilevers of two micropipettes were positioned parallel to the $xy$-plane, situated between the glass slides and submerged in the cell culture medium. Each micropipette was connected to a micropipette injection holder (IM-H3 Injection Holder Set by Narishige Lifemed Co., Ltd.), which was linked via plastic tubing (i.d.= 1 mm; Narishige, model no. CT-1) to syringes (sizes ranging from 10 to 20 mL) filled with water. The holders were securely mounted on optical bases (Dynamically Damped Post, 14" Long, Metric, Ø1.5", Thor Labs, model no. DP14A/M) fixed to an active vibration isolation stage (MVIS 30×30 model by Newport Corporation), minimizing interference from environmental vibrations or nearby movement. Suction pressure, controlled via the syringes, was used to enable the micropipettes to firmly grasp the clusters.

A straight micropipette was attached to a linear motor (Thorlabs Kinesis Brushed Motor Controller) to enable controlled motion along the $x$-axis, with an initial acceleration of 1 mm/s² and a constant velocity of 20 µm/s. An L-shaped micropipette was positioned such that the bent portion of its cantilever was parallel to the cantilever of the straight micropipette. The L-shaped micropipette remained stationary throughout the experiment and functioned as a force sensor.

***Experimental procedure***

T-cells were extracted and activated at the University of Helsinki and moved to Aalto University for mechanical analysis in 6-well cell culture plates in a Styrofoam box at 37ºC. All samples were analysed within 12 hours of activation. The prepared cell media contained a high density of single T-cells and a small number of aggregates. Since the cells and aggregates were denser than the surrounding medium, they settled onto the bottom slide of the experimental chamber. To allow for proper settling, the sample was left undisturbed for a few minutes before beginning the experiment. Once an aggregate had stabilized on the bottom slide, the straight micropipette was used to lift it from the surface and position it between the slides. Both micropipettes then grasped the aggregate, ensuring that neither of them made contact with the slides. The linear motor was activated, causing the straight micropipette to move backward and stretch the cluster. This induced deflection in the stationary L-shaped micropipette, which served as a force sensor. The applied force and cluster deformation were determined by analysing the recorded images using MATLAB.

In the cell culture medium, a limited number of aggregates were present, exhibiting variations in size and shape. For the experiments, aggregates with a somewhat uniform shape, structure, and density were selected. Experiments were deemed invalid, and their results were excluded if any of the following conditions occurred: aggregate rotation, vertical displacement of the straight micropipette, rupture of the aggregate at the onset of the test without any observed stretching or deformation, or the appearance of jumps in the stress-strain graph. These jumps in the measured force were attributed to fluctuations in suction pressure during the experiment, leading to inconsistent data.

***Analysis of measurements***

MATLAB was used to determine the initial dimensions of the aggregate by measuring the distance between designated points in the experimental frames. In particular, the initial radius ($R_0$) and length ($L_0$) of the aggregate were determined from the first frame. The initial length was measured as the distance between the openings of the two micropipettes. To ensure measurement accuracy, each measurement was repeated five times, with the mean and standard deviation calculated and reported for subsequent calculations. We also used MATLAB to determine the deflection of the L-shaped micropipette as a function of time, $x(t)$. This deflection was utilized to calculate the applied force as a function of time, $F(t)$. Additionally, the change in the length of the cluster as $\Delta L = x_s - x$, where $x_s = vt$ is the motion of the straight micropipette. The Young's modulus was determined by fitting a line to the first 10–30 data points in the stress-strain graph.

***Validation and replication***

To ensure consistency and accuracy, stiffness measurements were performed on 5 to 25 clusters from each prepared cell media sample.

### G. Statistical analysis

We performed Welch's t-test (MATLAB 'ttest2') to statistically validate the difference between (*i*) un-activated and activated T-cells with respect to expression levels of CD69 and CD25 (FIG. 1c–d) and (*ii*) Young's moduli of aggregates activated by different agents (FIG. 7).

## DATA AVAILABILITY

The datasets used for plotting all graphs in the paper and examples of raw data files are shared on Zenodo (*76*).

## ACKNOWLEDGMENTS

This research was funded by the Väisälä project grant RESOLVE by the Finnish Academy of Science Letters (M.B.), the Jane and Aatos Erkko Foundation grant ROOTS (M.B.), and by Research Council of Finland (S.C.F., H.H.), Liv och Hälsa foundation (S.C.F.) and Magnus Ehrnrooth foundation (S.C.F.). We thank Dr. Maja Vuckovac for her help in the lab.

## APPENDIX A

There is a mouse model which has a knock-in (KI) TTT/AAA mutation in the β2 integrin that prevents the binding of Kindlin 3 to the cytoplasmatic tail of the protein and impairs part of its function, especially under conditions of shear flow. However, we found that the mutation has no effect on T-cell aggregate stiffness (**Fig. A1**), which correlates well will previous experiments that it also does not impact T-cell activation under these conditions (*67*).

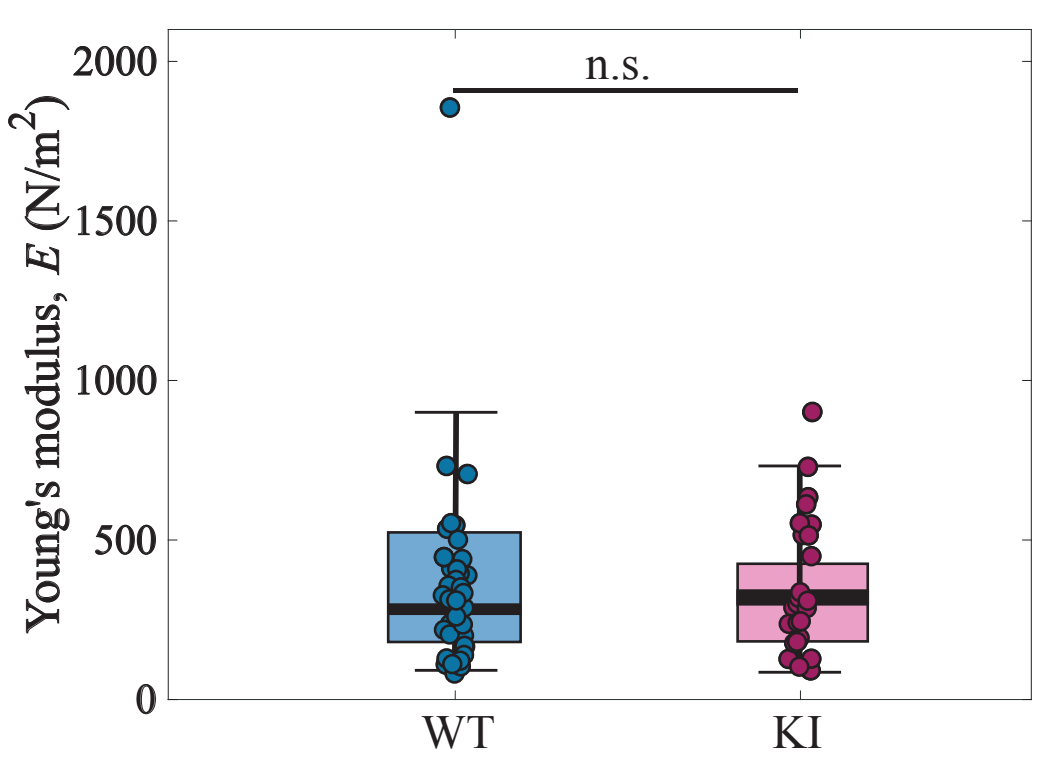

**FIG. A1.** Young's modulus of wild-type (WT) and knock-in (KI) T-cell aggregates 4 and 10h post-induction with PMA (10 ng/mL) and ionomycin (0.5 µg/ml). No significant difference was found between these two groups (n.s., $p > 0.05$).

## APPENDIX B

T-cell activation leads not only to cell aggregation and increased expression of activation markers, but also to changes in cellular metabolism. To investigate whether increased aggregate stiffness of W-7-treated cells was associated with an increased rate of cellular metabolism, we investigated protein translation rates of PMA+ionomycin-treated cells versus those treated also with W-7, by utilizing a puromycin-incorporation assay (*77*). Interestingly, we observed a minor (not significant) increase in cell metabolism of W-7-treated cells (**FIG. A2**), indicating that aggregate stiffness may correlate with altered cellular responses during activation.

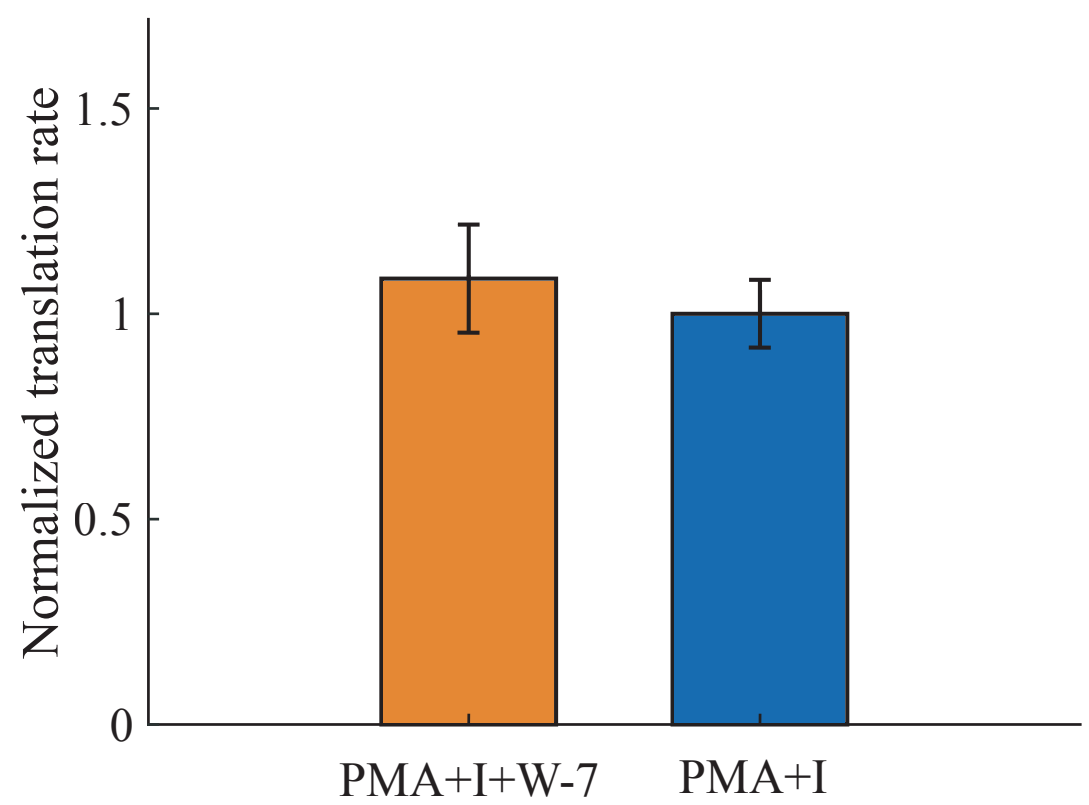

**FIG. A2.** Translation rate of T-cells at 24h post-induction with either PMA (10 ng/mL) and ionomycin (0.5 µg/ml) or with the same after a 30 min incubation with W-7 (50 µM). The results are normalized to the translation rate without W-7 (n=5). No significant difference was found between these two groups.